\documentclass[pra,longbibliography,amsfonts,amssymb,amsmath,floatfix,floats,a4paper]{revtex4-1}
\usepackage{graphicx}
\begin{document}

\title{Young's Double-Slit Experiment: ``What's Really Happening?"}

\author{N Gurappa}
\affiliation{Research Center of Physics, Vel Tech Multitech Dr.Rangarajan Dr.Sakunthala Engineering College, Avadi, Chennai, Tamil Nadu 600 062, India}

\begin{abstract}
A new wave-particle non-dualistic interpretation for the quantum formalism is presented by proving that the Schr\"odinger wave function is an `{\it instantaneous resonant spatial mode}' in which the quantum particle moves. The probabilities in quantum mechanics arise only for the observer's perspective due to the nature of doing experiments but they do not exist in  Nature. In other words, quantum mechanics itself is not a probabilistic theory. This view-point is proved by deriving the Born's rule by showing it to be equal to the experimentally observed relative frequencies. Also, it is shown that the classical and quantum mechanical times are one and the same. By identifying the already inherently existing mechanism for the collapse of wave function within the quantum formalism, the present interpretation naturally proves why the Copenhagen interpretation is very successful in explaining the experimental outcomes. The Young's double-slit experiment, the Wheeler's delayed choice experiment and the Afshar's experiment are unambiguously explained at a single quantum level. Both {\it measurement problem} and {\it retrocausality} are absent in the present non-dualistic interpretation of quantum mechanics.
\end{abstract}
\maketitle

 \section{Introduction}

Quantum formalism is the most successful theoretical description of Nature and it never imposes any limitations on its validity only to microscopic objects. It can, in principle, be applied to materials of any scale. Today we know that all the particles like photons, electrons, protons, atoms, molecules {\it e.t.c.}, exhibit the wave-particle duality \cite{spi1,spi2,spi3,spi4,spi6}. Prof. Feynmann said, "We choose to examine a phenomenon which is impossible, {\it absolutely} impossible, to explain in any classical way, and which has in it the heart of quantum mechanics. In reality, it contains the {\it only} mystery"; here, the `phenomenon' stands for the wave-particle duality of a single quantum in the Young's double-slit experiment \cite{Feyn1}. The Schr\"odinger wave function is assumed to give rise to this wave-particle duality because a single quantum event is always observed as a particle-like well-localized chunk and it becomes necessary to collect a large number of individual events in order to build an interference pattern such that the wave nature of a single quantum can be inferred \cite{spi1,spi2,spi3,spi4,spi6}. But, such a wave-particle duality naturally demands a mechanism by which the wave passing through the double-slit becomes a particle on the detector screen. But, according to the mainstream Copenhagen interpretation, such a mechanism seems to be unnecessary because it treats the wave function as unreal and considers the wave function as representing merely the probability of finding a particle at some location;  `collapse' exists naturally in any probabilistic description. However, it leads to a non-intuitive conclusion that the quantum particle is a {\it probability wave} until it is observed as a particle and clearly denies the pre-existing real world independent of observation. Therefore, the observers seem to play a very special role even though they are absent in the Hamiltonian describing the quantum state of a single particle. On the other hand, if the wave function is considered to be materialistically and physically real and also representing the `wave-particle duality', then it's a must to have a definite mechanism for the collapse of the wave function which may demand the modification of the Schr\"odinger equation \cite{GRW}, the very basic equation of quantum mechanics.

There are various interpretations for quantum formalism, like, the Copenhegan interpretation \cite{Au} as already mentioned, the Bhomian mechanics \cite{Bohm}, many-worlds interpretation \cite{Everette} etc. Most of the interpretations try to give a physical meaning to the wave function but at the same time maintain the notion of probability, instead of deriving the Born's rule to be indeed equal to the experimentally observed relative frequencies. The Bohmian mechanics provides a beautiful picture for quantum phenomena by removing the wave-particle duality but, only at the cost of introducing an epistemology-ontology duality, i.e., the wave function provides an epistemological probability distribution for the lack of knowledge about the initial position of a particle and at the same time plays an ontological role like a guiding wave for the same particle's motion. How some physical quantity can, at the same time, play a dual role as both epistemological and ontological is questionable. 

A scientific theory must be formulated with respect to Nature but not with respect to the observers or what can be observed. However, finding out such a theory without observations is impossible. At the best, the observations should serve as `anchor-points', such that they must help to pull out the actual reality present in the Nature. Then, such a theory can not be said to be metaphysical provided it is a unique one and capable of explaining even mutually exclusive physical phenomena. Once such a theory is formulated using what can be observed, then it becomes independent of observations; though this statement sounds like a paradox.

If the experimentally observed quantum phenomena can aptly be summarized by the Schr\"odinger equation, then the natural question to ask is "What could the Nature be like such that the quantum formalism correctly predicts the experimental outcomes?". The solution precisely lies in answering, `What's the physical reality of the Schr\"odinger's wave function and how it is related to the observed particles?'. In this paper, a picture of wave-particle non-duality underlying the quantum reality, an unique possible solution, is presented which is analogous to the situation of a moving test particle in a curved space-time of the general theory of relativity. Since, the present non-dualistic interpretation naturally provides the mechanism for the `wave function collapse' within the quantum formalism, it can be visualized as the best physical representation for the Copenhagen interpretation. At the same time, pictorially it has some resemblance to the Bohmian mechanics but without any additions like the guiding equation. Only the time-independent non-relativistic quantum mechanics is considered here, because, its interpretation naturally goes through time-dependent and relativistic cases. 

In Section-II, it is argued that the space in which Nature dwells is a complex vector space but not the Euclidean. In Section-III, it is shown that the inner-product of a state vector with its dual can be visualized as an interaction within the quantum formalism. In section-IV, a mathematical argument is presented to show that the Schr\"odinger's wave function can be realized as an {\it Instantaneous Resonant Spatial Mode}. Also, how to visualized the space around us as a complex vector space as demanded by quantum mechanics is explained. In Section-V, the absolute phase associated with a state vector is shown to be responsible for which eigenvalue of an observable will be realized in a given experiment. The Born's rule is derived in Section-VI. The situations of sequential measurements and `no quantum jump' are discussed in Sections VII and VIII, respectively. In Section-IX, the equality of classical and quantum mechanical times are shown. What's really happening in the Young's double-slit experiment, the Wheeler's delayed choice experiment and Afshar's experiment are explained in greater detail in Sections X and XI. Section-XII contains the conclusions and discussions.   

\section{`Empty Space' - according to Quantum Mechanics}

The Born's rule, interpreting the square of the norm of the wave function as probability, is experimentally successful quantum algorithm. Yet at the fundamental level, it's not really a    pleasing one. Nature can not ensure probability by repeated measurements much before the creation of the physical system and the measuring device i.e., before Nature came into existence. If the probabilities are assigned after the creation of the physical system, then it must be statistically derivable, at least in the sense of Einstein. It's not even possible to argue that both the system and its probability came together simultaneously into existence, because, then we must have a direct and further irreducible equation for the probability but not the Schr\"odinger's equation. 

Though the physical state of a quantum system is described by a complex vector in some `abstract' Hilbert space, it is relevant to explain the experimentally observed data via the  Born's rule. The experimental setup exists in the $R^3$ Euclidean space ($R^3$ES) but it still captures the information from the respective abstract Hilbert space. So, it's clear that there is a fundamental mismatch between the actual space in which the quantum particles really live and our intuition about the particle being present in $R^3$ES. Though the absolute space can be felt intuitively as nothingness, it's important to note that its true nature is unavailable to us independent of the material phenomena happening in it. 

In the Newtonian paradigm, a point is located in $R^3$ES by specifying its coordinates by a rigid measuring rod and hence attaching the property of rigidity to the space itself. Such a rigid but empty space provides an absolutely passive and unchanging `stage' for all the physical phenomena happening. But in Einstein's special theory of relativity, the material objects have an intimate connection with the space-time in such a way that their relative speed never exceeds the Cosmic speed limit in any inertial frame of reference. In general theory of relativity, space-time is directly related to the energy-momentum distribution and is dynamical. It can bend, stretch, twist and even ripple and dictates the particles' motion to lie along the geodesics. {\it It's very important to identify the actual space in which a physical phenomenon is happening. Otherwise, the reality starts appearing to be strange, weird and counter-intuitive and even retro-causal}.

Starting from the classic Stern-Gerlack experiment, all the quantum phenomena are actually found to take place in a `complex vector space' rather than the $R^3$ES, i.e., $R^3$ES is insufficient to describe the experimentally observed quantum phenomena. Then why the macroscopic objects, which are obviously composites of `quantum entities', appear to live in $R^3$ES? So, it's clear that the space around us is indeed `complex' in nature even though it is perceived  `effectively'  as $R^3$ES. Once the notion of Euclidean space is given up and the complex vector space is accepted as the actual space in which the Nature dwells, then all the quantum phenomena make sense exactly like classical mechanics. In other words, quantum mechanics is equivalent to classical mechanics but in the complex vector space. Quantum mechanics is reveling a profound and remarkable property of the space (or space-time) itself which is quite different from that of general relativity. Keeping this in mind, the following axioms are proposed.

\noindent {\underline{{\bf Axiom-1:}}} {\it `Empty space' is in general an infinite dimensional complex vector space of continuous (and also discrete) dimensions}. 

Let ${\bf S}$ be the set of elements denoting the empty space. The most crucial property of ${\bf S}$ is 
$$ {\bf S} = {\bf S} \otimes {\bf S} \quad ; {\bf S} \in {\bf S} $$
where, $\otimes$ stands for the direct-product and $\in$, for `is an element of'.

\noindent $\bullet$ Any quantum mechanical Hilbert space, ${\bf H}$, is always a subset i.e., ${\bf H} \subset {\bf S}$; note that here ${\bf H} = {\otimes}_{i = 1}^N {\bf H}_i$, where $N$ stands for the total number of particles.

\noindent \underline{{\bf Axiom-2:}} {\it A precise set of elementary particles in this infinite dimensional complex vector pace, {\bf S}, with well-defined properties and interactions among them results in the macroscopic manifestation of matter with respect to which the eigenvalues of position operator effectively form the $R^3$ES.}

All the vectors in ${\bf H}$ ($\subset {\bf S}$), when represented in the position basis, are super-imposed on top of each other and can independently co-exist in the same region of $R^3$ES spanned by the eigenvalues of the position operator. In other words, in a given region, any number of spatial modes can coexist which corresponds to the tensor-product of the state vectors. Classical waves, like ripples on a water surface, never behave this way and also this situation is very different from the classical Newtonian and Einsteinan spaces which are uniquely described by the position eigenvalues alone. More details regarding ${\bf S}$ will be reported elsewhere.

\section{Schr\"odinger wave function as an Instantaneous Resonant Spatial Mode}

In the following, a simple argument is presented to show that the Schr\"odinger wave function is an `instantaneous resonant spatial mode' (IRSM). Let's consider the de Broglie's case of a particle executing force-free motion in one-dimension. Its classical Hamiltonian, $H$, is given by 
\begin{equation}
H = \frac{p^2}{2 m} = E  \label{CH}
\end{equation}
where, $m$ is the mass, $p$ is the momentum and $E$ is the total energy of the particle. The Hamiltonian equations of motion are given by
\begin{equation}
\dot{p} = - \frac{\partial H}{\partial x} = 0 \,\,;\,\,\, \dot{x} = \frac{\partial H}{\partial p} = \frac{p}{m}   
\end{equation}
which yield the following solutions,
\begin{equation}
x(t) = \frac{p(0)}{m} t + x(0)  \,\,\,; \,\, {\rm and} \,\,\, p(t) = p(0)   \label{CS}
\end{equation}
where, $\dot{x}$ and $\dot{p}$ refers to the time, t, derivatives of $x$ and $p$ and $x(0)$ and $p(0)$ are constants of integration corresponding to the initial position and initial momentum of the particle.

It's well-known that the canonical commutation relation $[\hat{x} \,\,,\,\, \hat{p}] = i \hbar$ is the heart of quantum formalism; where, $\hat{x}$ is the position operator, $\hat{p}$ is the momentum operator, $i = \sqrt{-1}$ and $\hbar$ is the reduced Plank's constant. In fact, it captures the essence of the de Broglie's hypothesis which states that every moving particle is associated with a wave nature, with wavelength $ \lambda = h/p$; where, $h$ is the Plank's constant. That's precisely the reason why, by following the Dirac's prescription, one gets the Schr\"odinger wave equation from any classical Hamiltonian by a simple replacement of the classical commuting variables $x$ and $p$ by non-commuting operators $\hat{x}$ and $\hat{p}$, respectively.    

Now, instead of replacing $x$ and $p$ by $\hat{x}$ and $\hat{p}$ in Eq. (\ref{CH}), the same is done for the classical solutions in Eq. (\ref{CS}). 
\begin{equation}
\hat{x}(t) = \frac{t}{m} \hat{p}(0) + \hat{x}(0)  \,\,\,; \,\, {\rm and} \,\,\, \hat{p}(t) = \hat{p}(0)   \label{qem}
\end{equation}
One sees that the constants $x(0)$ and $p(0)$ have to be treated as operators such that the validity of the original position-momentum commutation relation remains unaffected, i.e.,  
\begin{equation}
[ \hat{x}(t) , \hat{p}(t)] = [ \hat{x}(0) , \hat{p}(0)] = i \hbar
\end{equation}

Note that Eq. (\ref{qem}) is nothing but Heisenberg's equations of motion. Therefore, in the position representation, the Hamiltonian operator can be written as 
\begin{equation}
\hat{H} = - \frac{{\hbar}^2}{2 m} \frac{\partial^2}{\partial x^2} = E
\end{equation}
or 
\begin{equation}
\hat{H} = - \frac{{\hbar}^2}{2 m} \frac{\partial^2}{\partial x(0)^2} = E
\end{equation}
such that
\begin{equation}
\hat{H} \psi(x(t)) = E \psi(x(t)) \,\,; {\rm and} \, \hat{H} \psi(x(0)) = E \psi(x(0)) 
\end{equation}
Both $\psi(x(t))$ and $\psi(x(0))$ describe the same physical situation for the energy eigenvalue $E$. 

Also, from Eq. (\ref{qem}), one has
\begin{equation}
[\hat{x}(0) , \hat{x}(t)] = i \hbar \frac{t}{m} \,\,; \, [\hat{p}(0) , \hat{p}(t)] = 0
\end{equation}
therefore,
\begin{equation}
\hat{x}(0) |x(0)> = x(0) |x(0)> \,\,;\,\,\hat{x}(t) |x(t)> = x(t) |x(t)> \label{eve}
\end{equation}
The position eigenvalues sets, $\{x(0)\}$ and $\{x(t)\}$,  of $\hat{x}(0)$ and $\hat{x}(t)$ span the same one-dimensional space. However, any given position eigenstate $|x(0)>$ is a linear superposition of all the eigenstates of $\hat{x}(t)$ because, $[\hat{x}(0) , \hat{x}(t)] \ne 0$ for any non-zero value of $t$ and vice versa. In other words, the position vector spaces of $\hat{x}(0)$ and $\hat{x}(t)$ are twisted with respect to each other though they give rise to the same set of position eigenvalues and hence the same one-dimensional space. 

Here, it's worth mentioning that a given quantum particle carries both position and momentum eigenvalues simultaneously. For example, in the Young's double-slit experiment, the fact that one observes the particle at some location in the detector screen is due to the fact that the particle arrives there due to the momentum it carries. But, it should be noted that the spaces in which the position and momentum state vectors live are twisted with respect to each other due to the commutation relation, $[\hat{x}, \hat{p}] = i \hbar$. This situation is entirely different from the classical one where the vector spaces of position and momentum are not at all twisted because $[\hat{x}, \hat{p}] = 0$ and the both the spaces are superimposed on top of each other and appears to be a single space. 

The wave function, $\psi$, can be considered as a function on a set of all possible initial values, $\{x(0)\}$, of the particle or as a function on a set of all possible values at any time, $t$, $\{x(t)\}$. But, for any dynamical process happening in any given space, the set of all possible initial values (or values at any later time) is the space itself. Thus, one can arrive at the conclusion that the schr\"odinger wave function is like a field on the original one-dimensional space. Therefore, once a particle of momentum $p$ appears, say at some position eigen value $x_p(0)$, then $\psi(x(0))$ appears instantaneously everywhere in the entire one-dimensional space. At later time $t$, the position eigen value of the particle changes from $x_p(0)$ to $x_p(t)$ and the corresponding instantaneous eigen mode is $\psi(x(t))$. Therefore, $\psi$ can be said to be an IRSM and the Plank's constant $h$ can be realized as a coupling parameter between the particle and its IRSM. Here, coupling means that the particle is actually free to move but always confined within its IRSM, which is like a deterministic constraint. The spatial modes can appear or disappear instantaneously everywhere but the particles moving in these modes are subjected to to obey the cosmic speed limit in accordance with the special theory of relativity. Further, this instantaneous nature is precisely responsible for the outcome of Wheeler's delayed choice experiment, Einstein's spooky action-at-a-distance in the case of two or more entangled particles and also entanglement swapping experiments \cite{Gurappa}. 

Now, consider the generic case: Let $\hat{r}$ be the position operator with eigenstates $|r>$ and eigenvalues $r$ so that the set of all possible $r$ values span the $R^3$ES. When a free particle of definite momentum eigenvalue `$p$' is created, then a resonant spatial mode `$< {\bf r}|\psi>$' appears instantaneously everywhere in the entire space such that the particle's motion is completely confined within this mode, where $|\psi> \in {\bf H} \subset {\bf S}$. This situation is exactly like the case in general relativity where a particle moves in a curved space-time. Since $|{\bf r}>$ and $< {\bf r}|\psi>$ are in one-to-one correspondence, without loss of generality, the state vector itself can be called as an IRSM.

The IRSM can be visualized as follows:
\begin{eqnarray}
|\psi> = \int d {\bf r} |{\bf r}><{\bf r}|\psi> 
\end{eqnarray}   
At every eigenvalue ${\bf r}$ of the position operator $\hat{\bf r}$, attach the complex vector $|{\bf r}><{\bf r}|\psi> $. That's the visualization of IRSM in which the given particle will be present at some eigenvalue ${\bf {r_p}}$, carrying the corresponding vector  $|{\bf {r_p}}><{\bf {r_p}}|\psi> $. This particular visualization shows that the the quantum mechanical Hilbert space is not `some abstract space' but it is the actual space around us. Note that the vectors associated with the particle, $|{\bf {r_p}}><{\bf {r_p}}|\psi> $, and any other spatial point, $|{\bf r}><{\bf r}|\psi> $, are identical in their structure. Hence, all these vectors at every position eigenvalue, ${\bf r}$, possess properties exactly like the particle itself and this may give rise to an illusion that the particle exists simultaneously in all possible positions before the measurement.

Since the IRSM obeys the Schr\"odinger equation, which is a partial differential equation, it's necessary to impose the boundary conditions. The initial boundary condition is a point in the complex space where the momentum is originated and remains unaltered as long as the particle sustains with the same momentum, i.e., `origin remains unaltered as long as the momentum of the particle gets unaffected' - is an important property of the complex vector space. This particular property is based on our common sense experience. For example, consider a star at a distance of some millions of light years away. When we look at it, we are not actually looking at what it is right now but how it is before some millions of years ago. Right now, anything might have happened like it might have exploded into a supernova or swallowed by a black-hole etc. We don't see all those details except the star as before a millions of years. It simply implies that the origins of photons are unaffected.  

The final boundary condition depends on where the particle will end up and need not be a fixed boundary condition (both boundary conditions are fixed for the case of bound systems). If the particle undergoes some momentum changing interaction, then the earlier IRSM disappears completely and a new IRSM corresponding to new momentum appears instantaneously with origin at the spatial point where the particle gained new momentum. In other words, the old origin disappears and a new origin appears at the same instant, independent of the distance of separation between them. {\it This particular picture of a particle flying in its own IRSM is non-dualistic in nature, further irreducible and is independent of any measurement procedure}. 

In the case when the particle is subjected to some potential, then the IRSM is a solution of the time-independent Schr\"odinger's wave equation with that potential and its behavior is exactly like a classical wave since it's a solution of a wave equation. 

\section{Inner-product as an interaction} 

This non-dualistic picture of a particle flying in its own IRSM is not analogous to any classical wave, though the IRSM obeys the Schr\"odinger wave equation. It's well-known that the
square of amplitude of a classical wave is proportional to its intensity. Therefore, such an intensity can't be claimed for the IRSM.

If the particle is going to end up, say for example on a detector screen, then a dual vector, $<\psi|$, is induced in that screen and interacts with the IRSM, $|\psi>$, as given by the inner-product, $<\psi|\psi>$. Then, that particle will be observed at some location in this region of inner-product. Note that, this inner-product interaction happens the moment the particle appears at the source. Instead of the detector screen, some surface of a distant planet or an eye of some creature etc., can also be considered.

In fact, this inner-product interaction can be found easily within the quantum formalism itself. Let the state $|\psi>$ encounters a screen and gets scattered into some other state $|\psi^\prime>$. This process can be described by associating an operator, $\hat{O} =|\psi^\prime><\psi|$, to the screen so that 
$$ \hat{O} |\psi> = <\psi|\psi> |\psi^\prime> $$
Therefore, if the scattered state is discarded or it is a null-state, then the particle must have interacted or got absorbed at some location in the region of inner-product given by $<\psi|\psi>$. 

Suppose, the screen itself is described by a quantum state $|M>$ so that when the particle interacts with the screen, an entangled state $|M^\prime>|\psi^\prime>$ results. This process can be described by associating an operator, 
$$\hat{Q} =  |M^\prime><M| \otimes |\psi^\prime><\psi|$$
Here, $|M^\prime> |\psi^\prime>$ is subjected to the conservation laws and the inner-product interaction is given by $<M|M> <\psi|\psi>$. This entanglement case  and also the case where the Hermitian operator $\hat{O} =|\psi^\prime><\psi| + |\psi><\psi^\prime|$ is associated with the detector screen will be considered separately elsewhere.  

{\it Precisely due to this inner-product interaction, we effectively feel the space to be} $R^3$ES.

If the detector states do not have complete basis to span $|\psi>$ and is associated with a projection operator $\hat{P}$, then the IRSM seen by the detector is $|\psi_D> = \hat{P} |\psi>$ and the dual vector excited in it is $<\psi_D|$. Therefore, the interaction region for the particle detection is $<\psi_D|\psi_D>$.

\section{Principle of minimum phase and quantum jump}

Let us consider a classical scenario of tossing a coin in $R^3$ES. Using the classical Newtonian mechanics, it is possible, in principle, to predict exactly whether head or tail will occur on a flat ground. If one is ignorant about some parameters involved in the dynamics of the coin, then probability can be invoked. 

Consider a normal vector, $\hat{\bf n}$, to the head surface passing through the center-of-mass of the coin and let $\theta$ be the angle between $\hat{\bf n}$ and any parallel vector, $\hat{\bf g}$, to the gravitational force. Just before landing, consider the position of the coin at a height $h \le r$ above the horizontal ground surface; where, $r$ is the radius of the coin. If $\theta$ lies between $ \pi/2 \le \theta \le 3\pi/2 $, then tail will be the out come. If $ -\pi/2 \le \theta \le \pi/2 $, then head will occur i.e., depending upon in which range $\theta$ lies, the coin is forced to jump into either head or tail state. During the outcome, tail or head, the vector $\hat{\bf n}$ will be either parallel or anti-parallel to $\hat{\bf g}$. If $\hat{\bf n}$ is replaced by electron's spin magnetic axis and $\hat{\bf g}$ by magnetic force direction in a Stern-Gerlac (SG) apparatus, then the resulting situation of the coin is exactly identical to that of an electron considered by Bell \cite{Bell}. Note that, here both the coin and the electron were considered to be in $R^3$ES. It can be easily seen that if the space is a complex-Euclidean instead of $R^3$ES, then $\theta$ corresponds to the phase angle between the respective complex vectors $\hat{\bf n}$ and $\hat{\bf g}$. 

Now, let's consider the limit $r \rightarrow 0$. When $r = 0$, the above argument will not hold to be true anymore because the range of $\theta$ splitting into two equally probable ones happens due to the non-zero value of $r$. But, if one still wants to retain the same result even when $r = 0$, i.e., the vector $\hat{\bf n}$ should be ether parallel or anti-parallel to $\hat{\bf g}$, then (i) when $r \rightarrow 0$, the coin should acquire $\hat{\bf n}$ as a preferred direction attached to it and (ii) the gravitational force in the flat $R^3$ES or the complex-Euclidean space should be considered in such a way that it gives raise to that result. Instead of the coin, let's consider an electron which is known to be zero-dimensional object and has a spin magnetic axis attached to it. In order to this spin axis to orient either parallel or anti-parallel to the external magnetic field direction, one can consider the modified magnetic force equation \cite{Bell} (but it is not supported by experiments). Instead of flat $R^3$ES and without modifying the force equation, let's consider the case where the space is complex and it itself splits into two components. Then a point particle with a spin axis attached to it, has to either get into one or the other component because it can't split itself. Now, the spin vector and hence the particle will be force to enter into any one of the spatial component which makes a minimum phase angle with respect to the spin vector. Also note that, in the case of the spatial mode splitting into two components, the particle need not be a point particle. In the following, this situation is considered in detail for an electron in the SG apparatus:

According to the present non-duaistic interpretation, an electron flies in its own IRSM. Let $|s>$ be the IRSM corresponding to the original spin state of an electron in the complex vector space (spatial dependence is suppressed for simplicity). Let the unit operator along the direction of magnetic force in the SG apparatus be $1_{\rm op} = |\uparrow><\uparrow| + |\downarrow><\downarrow|$. Therefore, the IRSM encountering the SG apparatus is 
$$|s> = |\uparrow><\uparrow|s> + |\downarrow><\downarrow|s>$$
As one can easily see, akin to the coin in $R^3$ES jumping into either head or tail depending upon $\theta$, the electron spin in the complex vector space jumps into either $|\uparrow>$ or $|\downarrow>$ depending upon which of the complex numbers $<\uparrow|s>$ and $<\downarrow|s>$ has a minimum phase. It should be noted that even though the electron jumps into, say $|\uparrow>$, the empty mode $|\downarrow>$ still survives until the detection of the electron. So, it is clear that the IRSM can be decomposed into various components but not the particle moving in it. In other words, only the IRSM can undergo superposition but not the particle flying in it.
%%%%%%%%%%%%%  TO BE CHECKED
Through experiment, one observes only the preexisting properties of the particles. However, the values of the observed properties may get altered due to the act of observation.  
%%%%%%%%%%%%%%%%
\section{Derivation of the Born's rule} 

When the IRSM, $|\psi>$, is decompose into eigenstates, $|a_i>$; $i = 1, 2, 3, \cdots$, of an observable $\hat{A}$, then the particle jumps from $|\psi>$ to one of the eigenstate, say $|a_p>$, such that the `phase' of the complex number $<a_p|\psi>$ is minimum when compared with all other phases made by the remaining eigenstates. Note that, all other empty eigenstates are still ontologically present though the particle itself is in the minimum phase eigenstate, $|a_p>$. If detection takes place, then the particle will be found in $|a_p>$ with an eigenvalue $a_p$ because all other empty modes do not contribute. This is indeed the `projection postulate', `the reduction of state vector' or `the collapse of the wave function' advocated in the Copenhagen interpretation. The IRSM is
\begin{eqnarray}
|\psi> = \sum_i |a_i><a_i|\psi>
\end{eqnarray}
and it interacts with the excited dual-mode, $<\psi|$, in the detector as
\begin{eqnarray}
<\psi|\psi> = \sum_i <\psi|a_i><a_i|\psi> \xrightarrow{\text{Observation}} |<a_p|\psi>|^2 
\end{eqnarray}

As it's well-known, the eigenvalues $a_p$ are the direct measurable but not $|<a_p|\psi>|^2 $ which can be computed by repeating the same experiment several times with the same kind of particle states. Each particle is represented by the same $|\psi>$ but with different initial phases i.e., $e^{i \phi} |\psi>$. So, when the IRSM is decomposed into eigenstates of $\hat{A}$, then each particle sees a minimum phase with some particular eigenstate $|a_i>$ depending upon $\phi$ i.e., the phase of the complex number $e^{i \phi} <a_i|\psi>$. Each value of $\phi$ gives raise to a different minimum phase with different eigenstate such that out of total number of particles with random phases, a fraction of them will be found in $|a_i>$. In the limit of total number of particles tending to infinity, one can see that the relative frequency coincides with $|<a_i|\psi>|^2 $ such that, 
\begin{eqnarray}  
<\psi|\psi> = \sum_i <\psi|a_i><a_i|\psi> = \sum_i |<a_i|\psi>|^2 =  1
\end{eqnarray}
which is the Born's rule.

Here, it's worth mentioning a quote by Dirac \cite{Dirac}, "{\it Question about what decides whether the photon is to go through or not and how it changes its direction of polarization when it goes through can not be investigated by experiment and should be regarded as outside the domain of science}". The present non-dualistic picture completely agrees with the first part of the statement, "...can not be investigated by experiment" - because $<\psi|\psi>$ will not give any experimentally detectable phase information, but disagrees with the last part, " ...and should be regarded as outside the domain of science".

Therefore, 
\begin{eqnarray}
{\rm Frequency} \,\,{\rm of} \,\, {\rm observation} = {\rm Probability}\,\, {\rm in}\,\, {\rm Quantum} \,\, {\rm Mechanics} =  {\rm Born's}\,\,{\rm rule} \nonumber
\end{eqnarray}
Therefore, the conclusion is that the quantum mechanics itself is not a probabilistic theory. Probability or the frequency of occurrence arises due to the nature of doing experiment.
A remark is that the probability interpretation actually leads to the measurement problem and is the actual obstacle preventing the visualization of physical reality. Once one attaches probability to some physical process and describes it in terms of probability, then such a physical process can not be explained independent of observation or measurement device. In the present non-dualistic picture, there is no probability in Nature. Nevertheless, it is not possible to predict the future event in the effective $R^3$ES which we perceive, because the actual quantum phenomenon is taking place in the complex space and the absolute phase of the IRSM is unavailable to experimental observation due to the inner-product interaction. So, we are forced to observe the frequency of outcomes after an infinite number of repeated, identical measurements.

\section{Sequential selective measurements}

Consider three sequential detectors A, B and C represented by observables $\hat{A}$, $\hat{B}$ and $\hat{C}$ \cite{Sakurai}. Let $|a_i>$, $|b_j>$ and $|c_k>$ be the  eigen vectors of the operators $\hat{A}$, $\hat{B}$ and $\hat{C}$ with eigenvalues $a_i$, $b_j$ and $c_k$, respectively; where $i, j, k = 1, 2, 3, \cdots$. Let the detectors $A$, $B$ and $C$ select some particular $|a^\prime_i>$, $b^\prime_j>$ and $|c^\prime_j>$ and rejects the rest. Let $|\psi>$ be the IRSM in which the quantum particle is flying. The first detector A has its own vectors space spanned by the eigenstates $|a_i>$. When $|\psi>$ is subjected to A, it gets resolved into various components as
\begin{eqnarray} 
|\psi> = \sum_i <a_i|\psi> |a_i>
\end{eqnarray}
in which only one component $|a^\prime_i>$ is allowed to come out and the rest are blocked. If the initial phase of $|\psi>$ is such that it makes a minimum phase with $|a^\prime_i>$, then the particle will be present in this mode and passes on to the next detector B.  When the mode $|{{\tilde{a}}_i}> (\equiv <a^\prime_i|\psi> |a^\prime_i>)$ encounters the detector B, then it gets resolved in B's space as 
\begin{equation}
|{\tilde{a}_i}> = \sum_j <b_j|{\tilde{a}}_i> |b_j>
\end{equation}
Now, B allows only $|b^\prime_j>$ by blocking the rest, hence $ <b^\prime_j|{\tilde{a}}_i> |b^\prime_j>$ will encounter C. This vector in C space is 
\begin{eqnarray}
<b^\prime_j|{{\tilde{a}}_i}> |b^\prime_j> = <b^\prime_j|{{\tilde{a}_i}}> \sum_k <c_k|b^\prime_j> |c_k> \,\,.
\end{eqnarray}
Now the detector C allows only $|c^\prime_k>$ to come out. There should be another detector D which measures the outcome from C. So, in response to the mode $<b^\prime_j|{{\tilde{a}_i}}> <c^\prime_k|b^\prime_j> |c^\prime_k>$, an excited dual mode $<{{\tilde{a}_i}}|b^\prime_j> <b^\prime_j|c^\prime_k> <c^\prime_k|$ in D interacts as $|<b^\prime_j|{\tilde{a}}_i>|^2 |<c^\prime_k|b^\prime_j>|^2$. If the initial phase of $|\psi>$ is such that the particle in it passes through all the detectors A, B and C, then it will be found at D. If one sends a large number of particles, all are described by the same $|\psi>$ but with different initial phases, then out of total number of particles, the fraction detected by D is given by 
\begin{eqnarray}
<{\tilde{a}}_i|{\tilde{a}}_i> \xrightarrow[\text{at D}]{\text{frequency of detection}}  |<b^\prime_j|{{\tilde{a}}_i}>|^2 |<c^\prime_k|b^\prime_j>|^2 \label{1R}
\end{eqnarray}

Suppose, the detector B allows all its modes. Then C will encounter the superposition of these modes, i.e., 
\begin{equation}
\sum_j <b_j|{\tilde{a}}_i> |b_j> = |{\tilde{a}}_i>
\end{equation}
In B space, though the particle is present in a particular mode $|b^\prime_j>$, all other empty modes do present ontologically and if unblocked, they will alter the outcome at C. So, one has 
\begin{equation}
|{\tilde{a}}_i> = \sum_k <c_k|{\tilde{a}}_i> |c_k>
\end{equation}
When C allows only one component $|c^\prime_k>$ to pass through, then D encounters a mode, $<c^\prime_k|{\tilde{a}}_i> |c^\prime_k>$. Then the excited dual mode $<{{\tilde{a}_i}}|c^\prime_k> <c^\prime_k|$ in D interacts as $|<{\tilde{a}}_i|c^\prime_k>|^2$. So, the frequency of observation at detector D is
\begin{eqnarray}
<{\tilde{a}}_i|{\tilde{a}}_i> \xrightarrow[\text{at D}]{\text{frequency of detection}} |<{\tilde{a}}_i|c^\prime_k>|^2
\end{eqnarray}
which is very different from Eq. (\ref{1R}). Therefore, ontological presence of an empty mode has a physically observable effect.

Therefore, if one regards $|<b^\prime_j|{{\tilde{a}}_i}>|^2$ as a probability for the particle to go through the $|b^\prime_j>$ route in B and $|<c^\prime_k|b^\prime_j>|^2$ as its probability of finding at $|c^\prime_k>$, then their product obeys the usual rule of probability multiplication. If probability is really in play here, then the total probability, $P(c_k^\prime)$, for the particle to arrive at $|c^\prime_k>$ through all possible $|b^\prime_j>$ is be given by
\begin{equation}
P(c^\prime_k) = \sum_j |<b^\prime_j|{{\tilde{a}}_i}>|^2 |<c^\prime_k|b^\prime_j>|^2 \label{pbc}
\end{equation} 
and it must be the same as without the detector B. But, in the absence of B, $|<{\tilde{a}}_i|c^\prime_k>|^2$ is the total probability of finding the particle at C and which is entirely different from Eq. (\ref{pbc}). This example is a clear demonstration for the absence of probability in quantum mechanics. As it's shown using the non-dualistic interpretation that only the frequency of observation arises at the detector when measurements are made on a large number of identical systems. It's wrong to infere the existence of probability for a single system in the absence of an instrument.

But in the case $\hat{A}$ and $\hat{B}$ commute, it can be easily verified that
\begin{eqnarray}
\sum_j |<b^\prime_j|{{\tilde{a}}_i}>|^2 |<c^\prime_k|b^\prime_j>|^2 &=& |<a_i^\prime|\psi>|^2 |<{a}^\prime_i|c^\prime_k>|^2 \nonumber\\
&=& |<b_i^\prime|\psi>|^2 |<{b}^\prime_i|c^\prime_k>|^2
\end{eqnarray} 
with $<a_i^\prime|b_j^\prime> = \delta_{ij}$ (here, $\delta_{ij}$ is the Kronecker delta), because any given eigenstate of $\hat{A}$ will have projection only along any one eigenstate of $\hat{B}$ and the remaining projections along all other eigenstates are zero. So, it doesn't matter whether one blocks or unblocks all the vectors with zero projection and also whether B is present or absent. But, in the case when $\hat{A}$ and $\hat{B}$ do not commute, then the entire vector space spanned by the eigenstates of $\hat{A}$ will have a non-zero twist with respect to the $\hat{B}$ space so that, any eigenvector in $\hat{A}$ space will have non-zero components along all eigenstates of $\hat{B}$ and vice versa.   

\section{Zero phase and no quantum jumps}

When the state $|\psi>$ representing the IRSM of a particle is decomposed into various orthogonal eigenstates of an operator with continuous eigenvalues, then the particle, without any quantum jump, will naturally enter into one of the eigenstate whose phase is exactly same as that of $|\psi>$.

As an example, consider the space spanned by the eigenvalues, ${\bf r}$, of the position operator $\hat{{\bf r}}$ with eigenstates $|{\bf r}> $; where, ${\bf r} = \{x,y,z\}$.
$$|\psi> = \int d{\bf r} |{\bf r}> <{\bf r}|\psi>$$
Then the particle will be present in a state $|{\bf {r_p}}> <{\bf {r_p}}|\psi>$ whose phase is exactly same as $|\psi>$. Therefore, upon observation,
\begin{eqnarray}  
<\psi|\psi> = \int d{\bf r} <\psi|{\bf r}> <{\bf r}|\psi> \rightarrow |<{\bf {r_p}}|\psi>|^2 
\end{eqnarray}
Even though the particle is present in $|{\bf {r_p}}> <{\bf {r_p}}|\psi>$ with an eigen value ${\bf r}_p$, it's easy to see that when it hits the screen, the excited dual mode is $<\psi|$ but not $<\psi|{\bf {r_p}}> <{\bf {r_p}}|$ because the detector is not projecting out this particular state, $|{\bf {r_p}}>$ . Therefore, the interaction is given by $|<{\bf {r_p}}|\psi>|^2$. 

Different absolute phases of a state vector will have different senses of directions in the complex vector space and hence they will result either in different position eigen values or the same position eigen value with different directions of particle propagation.

%%%%%%%%%%%%%%%%%%%%%%%%%%%%%%%%%%%
\section{Equality of classical time and quantum mechanical time} 
In the following, it will be shown that the time parameter entering in quantum mechanics is the same as the one in classical mechanics. Since, I did not explicitly consider the time-dependent Schr\"odinger's wave equation, the propagators are derived using the Heisenberg's equations of motion.

From Eqs.(\ref{qem}) and (\ref{eve}), one has
\begin{equation}
\hat{x}(t) |x(t)> = x(t) |x(t)>
\end{equation}
or
\begin{equation}
(\hat{x}(0) + \frac{t}{m} \hat{p}(0)) |x(t)> = x(t) |x(t)> 
\end{equation}
Introducing the unit operator, $\int dx(0) |x(0)><x(0)|$, in the position space at time $t=0$, one has 
\begin{equation}
\int dx(0) \left(\hat{x}(0) + \frac{t}{m} \hat{p}(0) \right) |x(0)><x(0)|x(t)> = x(t) \int dx(0) |x(0)><x(0)|x(t)> 
\end{equation}
which results in the following first order partial differential equation,
\begin{equation}
\left(- i \hbar \frac{t}{m} \frac{\partial}{\partial x(0)} + x(0) - x(t) \right) <x(0)|x(t)> = 0 
\end{equation}
whose solution can be found to be
\begin{equation}
<x(0)|x(t)> = \exp\left\{-\frac{i m}{2 \hbar t} [x^2(0) - 2 x(0) x(t) + \alpha] \right\}  
\end{equation}
where, $-\frac{i m}{2 \hbar t} \alpha$ is an integration constant. Similarly, by considering
\begin{equation}
\hat{x}(0) |x(0)> = x(0) |x(0)>
\end{equation}
and making use of the identity operator, $\int dx(t) |x(t)><x(t)|$, in the position basis at time $t$, one gets the following differential equation,  
\begin{equation}
\left(i \hbar \frac{t}{m} \frac{\partial}{\partial x(t)} + x(t) - x(0) \right) <x(t)|x(0)> = 0 
\end{equation}
whose solution is 
\begin{equation}
<x(t)|x(0)> = \exp\left\{\frac{i m}{2 \hbar t} [x^2(t) - 2 x(0) x(t) + \beta] \right\} 
\end{equation}
where, $\frac{i m}{2 \hbar t} \beta$ is an integration constant. Since, we know that $<x(t)|x(0)> = <x(0)|x(t)>^\star$, where $\star$ stands for complex conjugation, one has
\begin{equation}
x^2(t) - 2 x(0) x(t) + \beta = x^2(0) - 2 x(0) x(t) + \alpha^\star 
\end{equation}
which yields,
\begin{equation}
\beta = \sigma + x^2(0) \,\,;   \alpha^\star = \sigma + x^2(t)
\end{equation}
where, $\sigma$ is a constant. Therefore, one can write down
\begin{equation}
<x(t)|x(0)> = \exp\left\{\frac{i m}{2 \hbar t} [x^2(t) - 2 x(0) x(t) + x^2(0) + \sigma] \right\} 
\end{equation}
or
\begin{equation}
<x(t)|x(0)> = e^{\sigma^\prime} \exp\left\{\frac{i m}{2 \hbar t} \left(x(t) - x(0) \right)^2 \right\}
\end{equation}
with $\sigma^\prime = \frac{i m}{2 \hbar t} \sigma$. From the requirement that, 
\begin{eqnarray}
\lim_{t\to0}<x(t)|x(0)> = \delta(x(t) - x(0)) \label{limit}
\end{eqnarray}
one can infer $e^{\sigma^\prime} = \sqrt{\frac{m}{2 \pi i \hbar (t_2 - t_1)}}$, but it can also be shown elegantly by considering the unit operators in the position basis at time $t$ and at $t = 0$ as, 
\begin{eqnarray}
1 &=& \int d x(t) |x(t)><x(t)| \nonumber\\ 
&=& \iiint d x^\prime(0) d x^{\prime \prime}(0) d x(t) |x^\prime(0)><x^\prime(0)|x(t)><x(t)|x^{\prime \prime}(0)><x^{\prime \prime}(0)| \nonumber \\
&=& \iiint d x^\prime(0) d x^{\prime \prime}(0) d x(t) |x^\prime(0)> F(x(t),x^\prime(0),x^{\prime \prime}(0)) <x^{\prime \prime}(0)|
\end{eqnarray}
where,
\begin{equation}
F(x(t),x^\prime(0),x^{\prime \prime}) \equiv e^{(\sigma^\prime + {\sigma^\prime}^\star)} \exp\left\{\frac{i m}{\hbar t} \left[x^\prime(0) - x^{\prime \prime}(0) \right] x(t) + \frac{i m}{2 \hbar t} \left(x^\prime(0) - x^{\prime \prime}(0) \right)^2 \right\} 
\end{equation}
Such that, 
\begin{equation}
\int d x(t) F(x(t),x^\prime(0),x^{\prime \prime}) = e^{2 \sigma^\prime_r} \frac{2 \pi \hbar t}{m} \delta(x^\prime(0) - x^{\prime \prime}(0)) 
\end{equation}
which yields, 
\begin{equation}
e^{\sigma^\prime_r} = \sqrt{\frac{m}{2 \pi \hbar t}} 
\end{equation}
where, $\sigma^\prime_r = (\sigma^\prime + {\sigma^\prime}^\star)/2 = {\rm Re} \{\sigma^\prime\}$ is the real part of $\sigma^\prime$. Therefore, one has 
\begin{equation}
<x(t)|x(0)> = e^{\sigma^\prime_i}\sqrt{\frac{m}{2 \pi \hbar t}} \exp\left\{\frac{i m}{2 \hbar t} \left(x(t) - x(0) \right)^2 \right\} 
\end{equation}
where, $\sigma^\prime_i = (\sigma^\prime - {\sigma^\prime}^\star)/2 = {\rm Im} \{\sigma^\prime\}$ is the imaginary part of $\sigma^\prime$, which can be evaluated from the requirement given in eq. (\ref{limit}) as,
\begin{eqnarray}
\delta(x(t) - x(0)) = \lim_{t\to0}<x(t)|x(0)> &=& \lim_{t\to0} e^{\sigma^\prime_i}\sqrt{\frac{m}{2 \pi \hbar t}} \exp\left\{\frac{i m}{2 \hbar t} \left(x(t) - x(0) \right)^2 \right\} \nonumber \\
&=& e^{\sigma^\prime_i} i^{\frac{1}{2}} \delta(x(t) - x(0))
\end{eqnarray}
which results in $e^{\sigma^\prime_i} i^{\frac{1}{2}} = 1 $, such that,
\begin{equation}
<x(t)|x(0)> = \sqrt{\frac{m}{2 \pi i \hbar t}} \exp\left\{\frac{i m}{2 \hbar t} \left(x(t) - x(0) \right)^2 \right\} 
\end{equation}
Or, if one allows the time parameter to vary from $t_1$ to $t_2$ instead of $0$ to $t$, then the above equation can also be written as follows;
\begin{equation}
<x(t_2)|x(t_1)> =  \sqrt{\frac{m}{2 \pi i \hbar (t_2 - t_1)}}\exp\left\{\frac{i m}{2 \hbar (t_2 - t_1)} \left(x(t_2) - x(t_1) \right)^2 \right\} \label{Lag1}
\end{equation}
Now, the same above analysis is carried out for the simple harmonic oscillator. Consider its Hamiltonian, 
\begin{equation}
H = \frac{p^2}{2 m} + \frac{1}{2} m \omega^2 x^2
\end{equation}
which yields the Hamiltonian equations of motion,
\begin{eqnarray}
x(t) &=& \cos(\omega t) x(0) + \frac{\sin(\omega t)}{m \omega} p(0) \nonumber \\
p(t) &=& - m \omega \cos(\omega t) x(0) +  \sin(\omega t) p(0)
\end{eqnarray}
Considering the position and momentum variables in the above equations as operators,
\begin{eqnarray}
\hat{x}(t) &=& \cos(\omega t) \hat{x}(0) + \frac{\sin(\omega t)}{m \omega} \hat{p}(0) \nonumber \\
\hat{p}(t) &=& - m \omega \cos(\omega t) \hat{x}(0) +  \sin(\omega t) \hat{p}(0)
\end{eqnarray}
such that $[\hat{x}(t) , \hat{p}(t)] = [\hat{x}(0) , \hat{p}(0)] = i \hbar$, $[\hat{x}(0) , \hat{x}(t)] = \frac{i \hbar}{m \omega} \sin(\omega t)$ and $[\hat{p}(0) , \hat{p}(t)] = 
i \hbar m \omega \sin(\omega t)$. Now, $\hat{x}(t) |x(t)> = x(t) |x(t)> $ and $\hat{x}(0) |x(0)> = x(0) |x(0)> $ yield,
\begin{equation}
<x(t_2)|x(t_1)> = \sqrt{\frac{m}{2 \pi i \hbar \sin(t_2 - t_1)}} \exp\left\{\frac{i m \omega}{2 \hbar \sin(\omega(t_2 - t_1))} G(x(t_2),x(t_1),t_2,t_1) \label{Lag2}
\right\}
\end{equation}
where,
\begin{equation}
G(x(t_2),x(t_1),t_2,t_1) \equiv \left(x^2(t_2) + x^2(t_1) \right) \cos(\omega (t_2 - t_1)) - 2 x(t_2) x(t_1)
\end{equation}
When, $t_2 - t_1 = \Delta t \rightarrow 0 $, both Eq. (\ref{Lag1}) and (\ref{Lag2}) can be written as 
\begin{eqnarray}
\lim_{{t_2}\to{t_1}}<x(t_2)|x(t_1)> &=&  \sqrt{\frac{m}{2 \pi i \hbar \Delta t}}\exp \left\{\frac{i}{\hbar} \Delta t \left[ \frac{m}{2} \left( \frac{x(t_2) - x(t_1)}{\Delta t} \right)^2 - \frac{1}{2}\left[V(x(t_2)) + V(x(t_1))\right] \right] \right\} \nonumber\\
&=&  \sqrt{\frac{m}{2 \pi i \hbar \Delta t}} \exp \left\{\frac{i}{\hbar} \int_{t_1}^{t_2} dt L(\dot{x}(t), x (t)) \right\} \label{Lagrangian}
\end{eqnarray}

Now, let us consider the energy eigenstate,
\begin{eqnarray}
|\psi> = \int d x(0) |x(0)><x(0)|\psi> = \iint d x(0) d x(t) |x(0)><x(0)|x(t)><x(t)\psi> \label{Phase}
\end{eqnarray}
The particle will be present at some particular eigenstates $|x_p(0>$ whose phase ${\rm ph}\{<x_p(0)|\psi>\}$ at time $t = 0$ is same as ${\rm ph}\{|\psi>\}$. From Eq. (\ref{Phase}), one has
\begin{eqnarray}
{\rm ph}\{|\psi>\} = {\rm ph}\{<x_p(0)|\psi>\} &=& {\rm ph} \{<x_p(0)|x_p(t)><x_p(t)|\psi>\} \nonumber\\
&=& {\rm ph} \{<x_p(0)|x_p(t)>\} + {\rm ph} \{<x_p(t)|\psi>\} \label{phase}
\end{eqnarray}
where, ${\rm ph} \{<x_p(t)|\psi>\}$ is the phase of the particle state at $t$.
The phases of the particle states at $t=0$ and $t$ will be different, i.e.,  
\begin{eqnarray}
{\rm ph}\{<x_p(0)|\psi>\} \ne {\rm ph} \{<x_p(t)|\psi>\} 
\end{eqnarray}
but, any infinitesimal variation of phase at $t=0$ results in the corresponding variation of phase at $t$, i.e.,
\begin{eqnarray}
\delta \{{\rm ph}<x_p(0)|\psi>\} = \delta \{{\rm ph} <x_p(t)|\psi>\} 
\end{eqnarray}
which, from Eq. (\ref{phase}), implies that
\begin{eqnarray}
\delta \{{\rm ph} <x_p(0)|x_p(t)>\} = 0 \label{VP}
\end{eqnarray}
Applying Eq. (\ref{VP}) to Eq. (\ref{Lagrangian}) yields the classical least action principle,
\begin{eqnarray}
\delta \int_{t_1}^{t_2} dt L(\dot{x}_p(t), x_p(t)) = 0 \label{LAP}
\end{eqnarray} 
Therefore, the eigenvalues of particle state can always be inferred to lie on a classical path. This inference along with the inner-product interaction is responsible for the perception of the space to be Euclidean. Though the result in Eq. (\ref{LAP}) is shown here for the cases of free particle and harmonic oscillator, it is true for any general case with an arbitrary potential and will be reported elsewhere. The traces of particle trajectories photographed in the particle detectors like the Wilson chamber, are the direct evidences of this conclusion. 

Now, let us assume a `free particle' (say for simplicity, in one dimension along $x$-axis) described by the state vector $|\psi>$, moving inside a medium which continuously interact with the particle state, then,
\begin{eqnarray}
<\psi|\psi> = \eta \int dx |<x|\psi>|^2 = \eta \int dx |\psi(x)|^2 
\end{eqnarray}
where, where, $\eta$ is a constant characteristic to the medium and $<\psi|\psi>$, is the inner product between the original IRSM, $|\psi>$ and the dual, $<\psi|$ excited in the medium. This interaction, $<\psi|\psi>$, at a given point, $x$, can be treated as a potential in the Schr\"odinger equation,
\begin{eqnarray}
H = \frac{{p}^2}{2 m} + \eta |\psi(x)|^2
\end{eqnarray}
 Since, the Hamiltonian is Hermitian, one obtains the time-dependent non-linear Schr\"odinger wave equation \cite{NonSch} as
\begin{eqnarray}
- \frac{\hbar^2}{2 m} \frac {\partial^2 \psi} {\partial {x^2}} + \eta |\psi|^2 \psi = i \hbar \frac{{\partial} \psi}{\partial t}
\end{eqnarray}

\section{Young's Double-Slit Experiment: What's really happening?}

\begin{figure}[htbp]
\centerline{\includegraphics[width=0.6\textwidth]{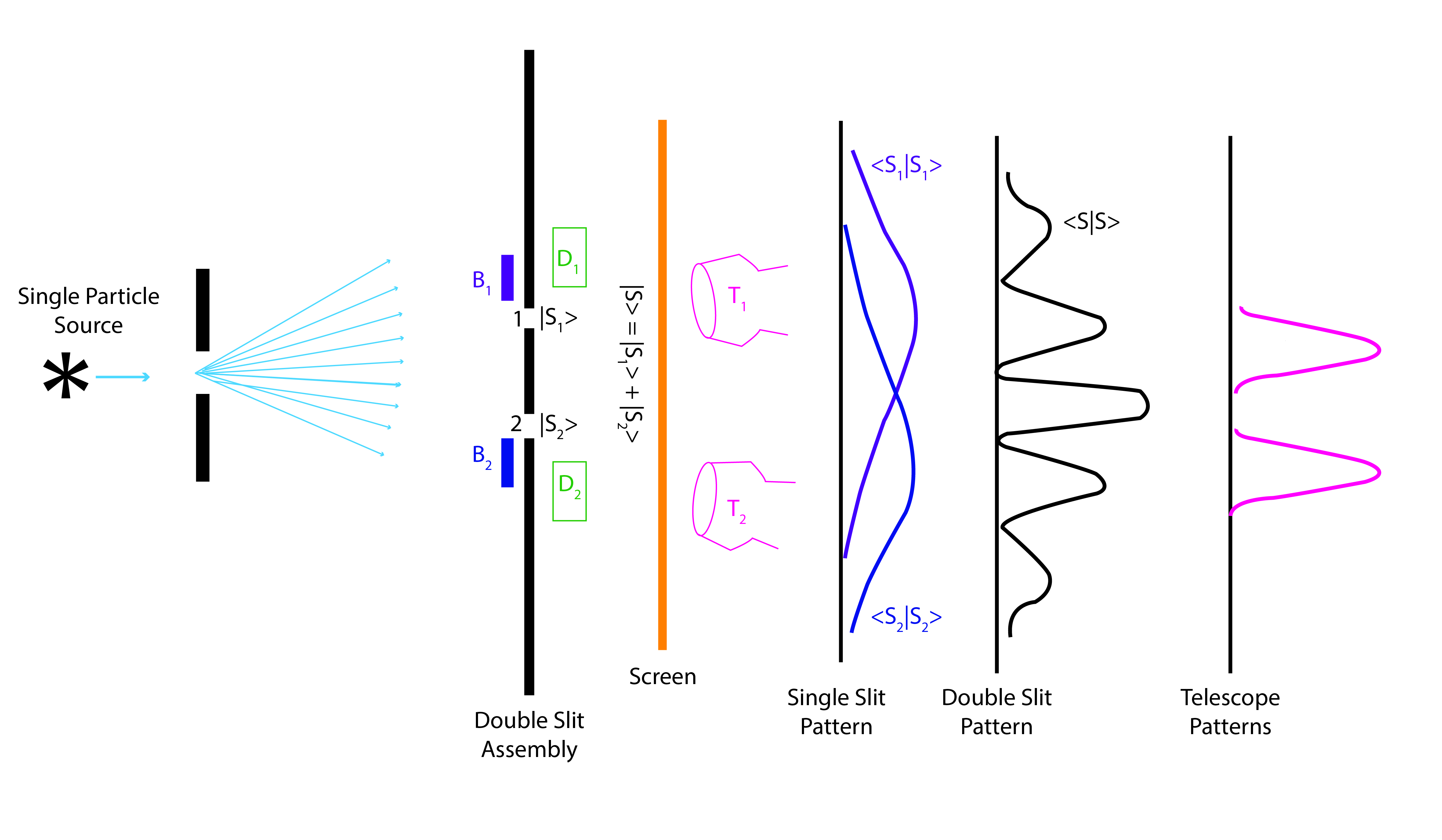}} \caption{\textbf{Diagram of single-particle Young's double-slit experiment}:~A source shoots single particles, one at a time, towards a double-slit assembly. $1$ and $2$ represents two slits through which the state vectors $|S_1>$ and $|S_2>$ emerge out and get superposed as $|S> = |S_1> +|S_2>$. $B_1$ and $B_2$ are two blockers which can block either slit $1$ or $2$ at any time. $D_1$ and $D_2$ are two detectors useful to find out through which slit any particle is passing to the screen. Immediately behind the screen, two telescopes, $T_1$ and $T_2$, are placed such that particles passing through $1$ and $2$ reach $T_1$ and $T_2$, respectively. After collecting a large number of individual particles, the observed particle distribution pattern on the screen and telescopes were also shown in the same diagram. When slit $2$ ($1$) is blocked, a distribution $<S_1|S_1>$ ($<S_2|S_2>$) is obtained. If both the slits are open, then the observed distribution is $<S|S>$.} \label{Figure1}
\end{figure}

Consider the Young's double-slit experiment as shown in Fig.~\ref{Figure1}. Here, the case of a single-photon shot onto the screen only after the registration of the previous photon is considered in order to elucidate the actual behavior of an individual quantum particle. 

The moment a photon appears at the source, its IRSM appears everywhere and hence through the double-slit assembly. Let $|S_0>$ be the IRSM of the photon created at the source. The projector, $\hat{P}_{\rm ds}$, associated with the double-slit assembly is
\begin{eqnarray}
\hat{P}_{\rm ds} = \int{d {\bf r}^{(1)}_1 |{\bf r}^{(1)}_1><{\bf r}^{(1)}_1|} + \int{d {\bf r}^{(2)}_2 |{\bf r}^{(2)}_2><{\bf r}^{(2)}_2|} \,\,,
\end{eqnarray}
where, $\{|{\bf r}^{(1)}_1>\}$ and $\{|{\bf r}^{(2)}_2>\}$ are position basis for slit $1$ and slit $2$, respectively. The state vector, $|S>$, emanating from double-slit to the screen is
\begin{eqnarray}
|S> = \hat{P}_{\rm ds}|S_0> &=& \int{d {\bf r}^{(1)}_1 |{\bf r}^{(1)}_1><{\bf r}^{(1)}_1|S_0>} + \int{d {\bf r}^{(2)}_2 |{\bf r}^{(2)}_2><{\bf r}^{(2)}_2|S_0>} \nonumber\\
&\equiv& |S_1> + |S_2> 
\end{eqnarray}
The dual-mode excited in the detector screen is $<S|$ which interacts with the IRSM according to the inner product $<S|S>$,
\begin{eqnarray}
<S|S> = <S_1|S_1> + <S_2|S_2> +  <S_1|S_2> + <S_2|S_1>
\end{eqnarray} 
Note that, this interaction has already formed on the screen instantaneously at the moment of the photon's creation at the source but remains unobservable until the hit of the photon at some position. The photon which is flying within the IRSM passes either through slit $1$ or $2$, depending on its initial phase. The moment photon's momentum changes either due absorption or scattering at the detector screen, then the entire IRSM disappears leaving the photon to contribute a point in the interaction region $<S|S>$. This is exactly the wave function collapse advocated in the Copenhagen interpretation but, without prescribing any  mechanism. Once the solution of the Schr\"odinger's wave equation is recognized as an IRSM, then the reduction of state-vector naturally exists within the quantum formalism itself.

The next photon appears at the source along with its IRSM whose phase will be different from the previous photon, i.e., its state may be given by $e^{i \phi} |S>$. The interaction region, $<S|S>$, is the same for all photons but their hits on the screen occur randomly at different locations due to different values of phases, $\phi$. The randomness in phases is due to its dependence on the detailed properties of the source and many other parameters, eventually like the particle's initial location in the entire Cosmos. After a large collection of photons landing at random positions, an interference pattern results on the screen which is nothing but the construction of the function $|<{\bf r}|S>|^2$ with individual points. 

It's worth recollecting a philosophical saying, "It is necessary for the very existence of science that the same conditions always produce the same result". Note that, according to the present non-dualistic interpretation, different absolute phases associated with different state vectors corresponding to the same physical situation result in different eigenvalues of an observable. Also, there is no way to prepare various initial states with the same absolute phase value. Therefore, the above saying seems to be in perfect agreement even in the context of quantum mechanical phenomena. 

If slit-1 (slit-2) is blocked, then a clump pattern corresponding to a single slit diffraction of slit-2 (slit-1), $<S_2|S_2>$ ($<S_1|S_1>$), appears on the screen. In the wave-particle dualistic interpretation, this single slit diffraction is attributed to the particle nature whereas, the double-slit diffraction i.e., the interference pattern, to the wave nature. But, in the present non-dualistic picture, the particle flying in its IRSM is independent of any experimental procedure and whether it's a single slit or a double-slit, the particle always move in the corresponding IRSM.

Classically, the photons were expected to leave a pattern of two strips on the screen as they pass one by one, some through slit-$1$ and some through slit-$2$, because they were thought to be traveling in the passive $R^3$ES. Nevertheless, an interference pattern reminiscent of wave nature appears suggesting that every individual photon is aware of whether one or both slits are open. Also, from the observed interference pattern an inference is made suggesting that a single particle-like photon `somehow' passes through both the slits simultaneously. In fact, trying to picturise the double-slit phenomenon in the passive $R^3$ES is the root for the concept of wave-particle duality. But, in the present non-dualistic interpretation, the photons are not actually moving in the $R^3$ES but in its own IRSM which obeys the Schr\"odinger equation. Also, it can easily be seen that the macroscopic objects naturally yield clump patterns matching our classical intuition because, their de Broglie wave length is extremely small when compared to the size of the object, the dimensions of slits and their separation.

Therefore, according to wave-particle non-duality, no surprise will occur when a detector observes through which slit a photon is really passing through. It should always appear as going through slit-$1$ or slit-$2$ like a particle but never through both the slits simultaneously like a wave. Now, the interference pattern disappears and two clump patterns appear as a proof for the observed particle behavior at the respective slits. Any momentum changing interaction of the photon with the detector's probe will result in the disappearance of the earlier IRSM, $|S>$, which had two origins at slit-1 and slit-2. A new IRSM (either $|S_1^\prime>$ or $|S_2^\prime>$) corresponding to new momentum appears whose single origin lies at the interaction point in the space and it will not pass through any slit in the direction of photon's motion. Its interaction with the detector screen is either $<S_1^\prime|S_1^\prime>$ or $<S_2^\prime|S_2^\prime>$. So, in the presence of detectors, clump patters occur and when the detectors are taken away, the interference comes back. This particular property viz, `{\it the disappearance of interference pattern and the appearance of clump patterns whenever the photons are watched through which slit they are going}', is an ultimate proof for the underlying particle nature of  photons (or any other material particles). Nature is providing a double-confirmation for the particle nature, i.e., first, when it was observed through which slit it was passing and the second, the disappearance of the interference pattern. If the underlying particle nature is absent, then the disappearance of interference pattern is impossible. Therefore, the so called quantum enigma, i.e., `when photons are watched, they appear to go through only one slit like particles, but, when they are not watched, then they seem to go through both the slits simultaneously like a wave' is an inference drawn due to the wave-particle duality by visualizing the quantum phenomenon in the passive $R^3$ES. 

Now, let's consider the Wheeler's delayed-choice situation\cite{Wheeler}. The screen is removed quickly exposing the twin telescopes, after a photon has already passed through the double slits. The interference pattern which would have occurred on the screen is lost and two clump patterns, one at each telescope, are formed. Using wave-particle duality, the inference drawn is that the photon retroactively rearranges its past history of passing through both the slits simultaneously like a wave to that of passing through only one slit at a time like a particle. Here, I would like to emphasized that the wave-particle duality implies retrocausality. 

If the screen is replaced by the twin telescopes while a photon is in mid-flight, then
\begin{eqnarray} 
(\hat{T}_1 + \hat{T}_2) |S> &=& (\hat{T}_1 + \hat{T}_2) (|S_1> + |S_2>) =  \hat{T}_1 |S_1> + \hat{T}_2 |S_2> \nonumber \\
&=& |\tilde{S}_1> + |\tilde{S}_2> \equiv |\tilde{S}>
\end{eqnarray}
where, $\hat{T}_1$ and $\hat{T}_2$ are operators associated with the telescopes, $T_1$ and $T_2$, since they are lens systems. Now, the old IRSM, $|S>$ is replaced to new IRSM $|\tilde{S}>$ but their origins remain unchanged, i.e., they are same for both $|S>$ and $|\tilde{S}>$. From whatever be the position of the photon during the replacement of $|S>$ to $|\tilde{S}>$, it continues to fly from there into the new IRSM, $|\tilde{S}>$. {\it The photon's motion is always continues even though its IRSM changes suddenly}. The continuity in photon's motion is governed by its conserved properties. Therefore, the observed photon distribution at telescopes is given by
\begin{eqnarray}
<\tilde{S}|\tilde{S}> = <\tilde{S}_1|\tilde{S}_1> + <\tilde{S}_2|\tilde{S}_2> 
\end{eqnarray}
which is a clump-pattern. Here, $<\tilde{S}_1|\tilde{S}_2> = 0$. 
Note that, in this present non-dualistic explanation, {\it the causality is preserved}, i.e., there is no retro-casual influences. 

Now, consider the case of the above Young's Double-Slit Experiment with Polarization filters $P_1$ and $P_2$ in the place of blockers $B_1$ and $B_2$, respectively. Then, the IRSM is given by
\begin{eqnarray}
|S>> = |S_1> |P_1> + |S_2> |P_2> 
\end{eqnarray}
and the interaction of IRSM on the screen is 
%%%%%%%%%%%%% eqn-modify
\begin{eqnarray}
<<S|S>>  = &&<S_1|S_1> <P_1|P_1> + <S_2|S_2> <P_2|P_2> \nonumber\\
&+&  <S_1|S_2> <P_1|P_2> + <S_2|S_1> <P_2|P_1>
\end{eqnarray}

Therefore, if $|P_1>$ and $|P_2>$ are orthogonal, then the interference vanishes.

Let $|P_1> = |H>$ and $|P_2> = |V>$. Then $|\psi>> = |S_1> |H> + |S_2> |V>$. Let us introduce a $45^o$ polarization rotator just before the screen. Then one has
\begin{eqnarray}
|S>> \rightarrow |\bar{S}>> = |{\bar{S}}_1> |\bar{H}> + |{\bar{S}}_2> |\bar{V}>
\end{eqnarray}
where, $|\bar{H}> = {1}/{\sqrt{2}} (|H> + |V>)$ and  $|\bar{V}> = {1}/{\sqrt{2}} (- |H> + |V>)$ , then on the screen one gets
\begin{eqnarray}
<<\bar{S}|\bar{S}>> &=& \frac{1}{2}  (<{\bar{S}}_1|{\bar{S}}_1> + <{\bar{S}}_2|{\bar{S}}_2> - <{\bar{S}}_1|{\bar{S}}_2> - <{\bar{S}}_1|{\bar{S}}_2>) <H|H> \nonumber \\ 
&+& \frac{1}{2} ( <{\bar{S}}_1|\bar{S}_1> + <{\bar{S}}_2|\bar{S}_2> + <{\bar{S}}_1|{\bar{S}}_2>  + <{\bar{S}}_1|{\bar{S}}_2>) <V|V> \nonumber \\
&=&  <{\bar{S}}_1|{\bar{S}}_1> + <{\bar{S}}_2|\bar{S}_2>
\end{eqnarray}
Suppose, a Wollaston prism with unit operator $1_{\rm op} = |H><H| + |V><V|$ is introduced in between polarization rotator and screen, then it will resolve $|\bar{S}>>$ into two orthogonal components $|H>$ and $|V>$, which can be detected by two independent detectors $D_H$ and $D_V$; obviously the outputs are given by
\begin{eqnarray}
{D_H}_{\rm output} = \frac{1}{2}  (<{\bar{S}}_1|{\bar{S}}_1> + <{\bar{S}}_2|{\bar{S}}_2> - <{\bar{S}}_1|{\bar{S}}_2> - <{\bar{S}}_1|{\bar{S}}_2>)
\end{eqnarray}
and 
\begin{eqnarray}
{D_V}_{\rm output} = \frac{1}{2} ( <{\bar{S}}_1|{\bar{S}}_1> + <{\bar{S}}_2|{\bar{S}}_2> + <{\bar{S}}_1|{\bar{S}}_2>  + <{\bar{S}}_1|{\bar{S}}_2>) 
\end{eqnarray}
Note that, the polarization rotator can be randomly introduced or removed before a photon actually passes through it. As it was explained above in the Wheeler's delayed choice experiment, the photon continues to fly even during the IRSMs are being changed randomly. This method was used in an experiment by Jacques et al. \cite{Jacques}, which uses Mach-Zehnder interferometer instead of Young's double-slit assembly. Similar experiment was also done using single atoms \cite{SingleAtom}.

\section{Afshar's experiment and Bohr's complementarity }

According to the wave-particle duality, it's the same object which can be observed like a particle but propagates like a wave in the absence of observation. Even though the given object possesses both the particle and wave natures, only one nature becomes possible to observe at a given time. Which one of the natures is observed depends upon the type of measuring device used. Here, the particle and wave natures are complementary behaviors meaning that both can't be observed simultaneously in the same experimental setting. This is the essential idea of Bohr's principle of complementarity (PC)\cite{Bohr1,Bohr2,Wheeler}. Here, I would like to emphasize the case of a single slit experiment where one observes a diffraction pattern which is due to wave nature. Nevertheless, on the detector screen, one always observes particles landing as a well localized chunks. Therefore, the observation is in terms of particle nature and the overall observed phenomenon, after collecting a large number of particles, reflects a wave nature. So, both particle and wave natures along with `which slit information' are completely available, because there is only one slit. This is clearly against the PC. Further, if one tries to detect whether the particle is really passing through the slit or not, then the diffraction pattern disappears. Therefore, in the absence of observation, one can always infer the particle as going through a particular slit and hence, which slit information does not account for the particle behavior. Also, when the `which hole detector' is turned on, the observed intensity on the screen is that of a spherical wave whose origin lies in the region where the detector probe interacts with the particle in the vicinity of the slit. The same is observed as a overlap of two spherical wave intensities in the case of double-slits, which can be confirmed by observing the Hanbury-Brown-Twiss effect \cite{HBT1,HBT2}. 

In the wave-particle duality, both the particle and wave natures are classical concepts stitched together and when these two natures are treated as complementary to each other, then one arrives at the PC. It should be noted that by using the quantum formalism, no physical mechanisms are provided for stitching both the natures together and also how one nature makes instantaneous transition to the other nature. 

The Afshar's experiment\cite{Afshar1,Afshar2} as shown in Fig. \ref{Figure3}, is a variant of the Young's double-slit experiment to verify the validity of the PC. A laser beam is made to fall on the pinholes 1 and 2. The state vectors, $|S_1>$ and $|S_2>$, emanating from the pinholes 1 and 2 undergo superposition and form an interference pattern on a vertical screen, if present. A vertical grid of thin wires of equal thickness, $\Delta x_i = \Delta x \,\,;\,i = 1,2,3 \cdots $, is placed in the place of the screen so that the wires lie exactly in the regions of destructive interference. A convex lens is placed away from the dual-pinholes but just after the grid so that the images of the pinholes 1 and 2 fall on the photon-detectors $D_1$ and $D_2$, respectively. 

By blocking the pinhole-2 (pinhole-1), the light intensity due to the image of pinhole-1 (pinhole-2) is registered by $D_1$ ($D_2$) both in the presence and absence of the grid.
By comparison, the intensity in the presence of the grid is found to be measurably less than the intensity in the absence of the grid. When both the pinholes were kept open, then the total photon flux through the dual pinholes were found to be almost the same as that one detected by both $D_1$ and $D_2$. It implies that the presence of the grid had no appreciable effect like an obstacle for the photon flux which in turn suggests to infer the existence of an interference pattern at the grid because the thin wires were placed at the dark fringes. Note that this is mere an inference because the total interference pattern was not actually recorded. 

Therefore, if the interference pattern is truly present, then it implies that the photons behaved like a wave. But the same photons gave rise to the images of pinholes at $D_1$ and $D_2$  due to momentum conservation which lead to conclude that each photon behaved like particle. If one uses wave-particle duality, this situation seems to be paradoxical because a given photon has to pass through both the pinholes to form an interference at the grid location but at the same time it has to pass through any one of the pinhole to behave like a particle at either $D_1$ or $D_2$, in the same given experimental arrangement which is clearly against the PC. Here, I would like to emphasize that the inference about a single photon simultaneously going through both the slits to produce interference itself violates the law of conservation of momentum and is a fundamental drawback in the concept of wave-particle duality which is also the cause for believing the Nature to be intrinsically random, probabilistic and also retrocasual. Wave-particle duality does not contain any paradox but it itself is a paradox. 

Now, I consider the results and conclusions of the same experiment for explaining within the present non-dualistic interpretation of quantum formalism. According to the non-duality, every quantum object moves in its own IRSM which is clearly against the PC and hence Afshar's conclusion is immediately supported. The results of Afshar's experiment is analyzed in the following:
 
\begin{figure}[htbp]
\centerline{\includegraphics[width=0.4\textwidth]{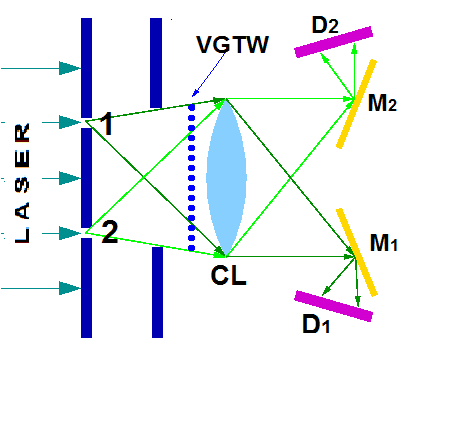}} \caption{\textbf{Diagram of Afshar experiment}:~ A laser beam, passing through two closely spaced circular pinholes 1 and 2, is refocused by a convex lens, CL, such that the images of pinhole-1 and pinhole-2 fall on the photon-detectors $D_1$ and $D_2$, respectively. When pinhole-2 (pinhole-1) is blocked, then a photon passing through pinhole-1 (pinhole-2) is detected by $D_1$ ($D_2$). A vertical grid of thin wires, VGTW, is placed just before the CL, in the region of interference due to the pinholes 1 and 2 so that the wires lie in the dark fringes. $M_1$ and $M_2$ are two totally reflecting mirrors.} \label{Figure3}
\end{figure}

Consider the case of single photons shot at the pinholes 1 and 2 so that any photon is fired only after the registration of the previous photon by any one of the photon-detectors $D_1$ and $D_2$. The photon state, $|S>$, is a superposition of states emanating from the dual pinholes, i.e.,
\begin{equation}
|S> = |S_1> + |S_2>
\end{equation}
where, $|S_1>$ and $|S_2>$ are from pinholes 1 and 2, respectively. The projector, $\hat{P}_g$, associated with the vertical grid of thin wires is given by 
\begin{equation}
\hat{P}_g = \sum_i \int_{x_i}^{x_i + \Delta x_i} d x_i^\prime |x_i^\prime><x_i^\prime| 
\end{equation}
Since the sets of position eigen values $[x_i , X_i + \Delta x_i]$, representing the thickness of the thin wires (here, [ , ] stands for a closed set, but not a commutator); $i = 1,2,3, \cdots$, lie in the dark fringes, the inner-product interaction at the grid surface is given by
\begin{eqnarray}
<S|\hat{P}_{g}^\dagger \hat{P}_g|S> = <S|\hat{P}_{g}|S> = & & <S_1|\hat{P}_{g}|S_1> +  <S_2|\hat{P}_{g}|S_2> \nonumber \\ 
&+& <S_1|\hat{P}_{g}|S_2> + <S_2|\hat{P}_{g}|S_1> \nonumber \\
\approx & & 0 \label{df}
\end{eqnarray}
where, $\hat{P}_{g}^\dagger = \hat{P}_g$,  $\hat{P}_g^2 = \hat{P}_g$ and $<S|\hat{P}_{g}^\dagger$ is the excited dual state in the grid. Therefore, one has from Eq. (\ref{df}),
\begin{eqnarray}
<S_1|\hat{P}_{g}|S_1> +  <S_2|\hat{P}_{g}|S_2> \approx - <S_1|\hat{P}_{g}|S_2> - <S_2|\hat{P}_{g}|S_1> \label{cancel} 
\end{eqnarray} 
We know from the Young's double-slit experiment that $<S_1|\hat{P}_{g}|S_1>$ and $<S_2|\hat{P}_{g}|S_2>$ are not independently equal to zero when only either pinhole-1 or pinhole-2 is opened but their sum can be exactly canceled by the term in the R.H.S of the Eq. (\ref{cancel}), by opening both the slits. Therefore, it's possible to choose the thickness of the thin wires sufficiently small so that the Eq. (\ref{df}) is satisfied when both the pinholes are opened and at the same time both $<S_1|\hat{P}_{g}|S_1>$ and $<S_2|\hat{P}_{g}|S_2>$ will appreciably reduce the observed frequencies, $<S_1|S_1>$ and $<S_2|S_2>$, at $D_1$ and $D_2$, when either pinhole-1 or pinhole-2 is kept open, respectively; here, $<S_1|S_1>$ and $<S_2|S_2>$ corresponds to the case without the grid. Therefore, the results of the Afshar's experiment can be explained within the non-dualistic interpretation of quantum formalism according to which a photon flies in its own IRSM, $|S>$, and hence naturally avoids the regions of dark fringes. Also, any given photon passes through either pinhole 1 or 2 at a given moment and due to momentum conservation, it will be detected either at $D_1$ or $D_2$, respectively. Therefore, it's true that in this single experimental setup, one can have both the interference pattern at the locations of the thin wires but a superposed state in the spaces between the wires and the information about through which pinhole the photon actually passed through.

\section{Conclusions and Discussions}

In conclusion, I have given a mathematical argument to show that the Schr\"odinger's wave function or equivalently the state vector of a quantum particle is an Instantaneous Resonant Spatial Mode (IRSM) in which the quantum particle flies. This situation is referred as wave-particle non-duality, since the IRSM and its particle always appear together as a single entity. This picture is analogues to the objects moving in the curved space-time of general theory of relativity. Once the underlying space is accepted as a complex vector space rather than the Euclidean, then the quantum phenomena exactly resemble the classical ones. The Euclidean space is perceived only `effectively' due to inner-product interaction and also due to the equality of classical and quantum mechanical times. Moreover, in this complex vector space, Quantum mechanics becomes a fully deterministic theory at a single quantum level and not at all a  probabilistic theory. The Born's rule arises due to the nature of doing experiment and is shown to be equal to the experimentally observed relative frequencies. This non-dualistic interpretation is the only interpretation from which a non-linear Schr\"odinger's wave equation can be derived. The Young's double-slit experiment, Wheeler's delayed choice experiment and Afshar's experiment are unambiguously explained. Particularly, the results of Afshar's experiment was in direct agreement with wave-particle non-duality. It was also shown that how the Copenhagen interpretation is naturally contained in the non-dualistic interpretation. The present interpretation overrules both the measurement problem and retrocausality, since they do not exist in the Nature. The present work reported here is a confirmation of what Einstein said, " God does not play dies". Since the present interpretation is a visualization of nature of reality reflected within the quantum formalism, it will go through both time-dependent and relativistic quantum mechanics. In the relativistic case, the IRSM is such that, apart from obeying the usual quantum mechanical commutation relations, it takes care of the cosmic speed limit for its resonant particle, though it itself can change instantaneously. Without much difficulty, it can be seen that the registered physical phenomena are independent of relative frame of reference. I conclude by saying, "Nature does not play dies in order to run our casual quantum mechanical Universe".

%references

 \end{document}